\documentclass[twocolumn]{aastex62}
\pdfoutput=1 
\usepackage{amsmath,amstext}
\usepackage[T1]{fontenc}
\usepackage[figure,figure*]{hypcap}
\usepackage{microtype} 

\newcommand{\kms}{km~s$^{-1}$}
\newcommand{\tdep}{$\tau_\textrm{dep}$}

\shorttitle{Star formation efficiency in NGC 628}
\shortauthors{Kreckel et al.}

\begin{document}

\title{A 50 pc scale view of star formation efficiency across NGC 628}

\author{K.~Kreckel}
\affiliation{Max Planck Institut f\"{u}r Astronomie, K\"{o}nigstuhl 17, D-69117 Heidelberg, Germany}
\author{C.~Faesi}
\affiliation{Max Planck Institut f\"{u}r Astronomie, K\"{o}nigstuhl 17, D-69117 Heidelberg, Germany}
\author{J.~M.~D.~Kruijssen}
\affiliation{\textls[-10]{Astronomisches Rechen-Institut, Zentrum f\"{u}r Astronomie der Universit\"{a}t Heidelberg, M\"{o}nchhofstra{\ss}e 12-14, D-69120 Heidelberg, Germany}}
\author{A.~Schruba}
\affiliation{Max-Planck-Institut f\"{u}r extraterrestrische Physik, Giessenbachstrasse 1, D-85748 Garching, Germany}
\author{B.~Groves}
\affiliation{Research School of Astronomy and Astrophysics, Australian National University, Weston Creek 2611, Australia}
\author{A.~K.~Leroy}
\affiliation{Department of Astronomy, The Ohio State University, 140 West 18th Ave, Columbus, OH 43210, USA}

\author{F.~Bigiel}
\affiliation{\textls[-12]{Institute f\"{u}r theoretische Astrophysik, Zentrum f\"{u}r Astronomie der Universit\"{a}t Heidelberg, Albert-Ueberle Str. 2, D-69120 Heidelberg, Germany}}

\author{G.~A.~Blanc}
\affiliation{The Observatories of the Carnegie Institution for Science, 813 Santa Barbara Street, Pasadena, CA 91101, USA}
\affiliation{Departamento de Astronom\'ia, Universidad de Chile, Casilla 36-D, Santiago, Chile}

\author{M.~Chevance}
\affiliation{\textls[-10]{Astronomisches Rechen-Institut, Zentrum f\"{u}r Astronomie der Universit\"{a}t Heidelberg, M\"{o}nchhofstra{\ss}e 12-14, D-69120 Heidelberg, Germany}}

\author{C.~Herrera}
\affiliation{IRAM, 300 rue de la Piscine, 38406 Saint Martin d'H\'eres, France}

\author{A.~Hughes}
\affiliation{CNRS, IRAP, 9 Av. du Colonel Roche, BP 44346, F-31028 Toulouse cedex 4, France}
\affiliation{Universit\'{e} de Toulouse, UPS-OMP, IRAP, F-31028 Toulouse cedex 4, France}

\author{R.~McElroy}
\affiliation{Max Planck Institut f\"{u}r Astronomie, K\"{o}nigstuhl 17, D-69117 Heidelberg, Germany}

\author{J.~Pety}
\affiliation{IRAM, 300 rue de la Piscine, 38406 Saint Martin d'H\'eres, France}
\affiliation{Sorbonne Universit\'e, Observatoire de Paris, Universit\'e PSL, \'Ecole normale sup\'erieure, CNRS, LERMA, F-75005, Paris, France}

\author{M.~Querejeta}
\affiliation{European Southern Observatory, Karl-Schwarzschild-Stra{\ss}e 2, D-85748 Garching bei M\"{u}nchen, Germany}
\affiliation{Observatorio Astron{\'o}mico Nacional (IGN), C/Alfonso XII 3, Madrid E-28014, Spain}

\author{E.~Rosolowsky}
\affiliation{4-183 CCIS, University of Alberta, Edmonton, Alberta, Canada}

\author{E.~Schinnerer}
\affiliation{Max Planck Institut f\"{u}r Astronomie, K\"{o}nigstuhl 17, D-69117 Heidelberg, Germany}

\author{J.~Sun}
\affiliation{Department of Astronomy, The Ohio State University, 140 West 18th Ave, Columbus, OH 43210, USA}

\author{A.~Usero}
\affiliation{Observatorio Astron{\'o}mico Nacional (IGN), C/Alfonso XII 3, Madrid E-28014, Spain}

\author{D.~Utomo}
\affiliation{Department of Astronomy, The Ohio State University, 140 West 18th Ave, Columbus, OH 43210, USA}

\correspondingauthor{K. Kreckel}
\email{kreckel@mpia.de}

\begin{abstract}
Star formation is a multi-scale process that requires tracing cloud formation and stellar feedback within the local ($\lesssim$\,kpc) and global galaxy environment.  
We present first results from two large observing programs on ALMA and VLT/MUSE, mapping cloud scales ($1\arcsec = 47$~pc) in both molecular gas and star forming tracers across $90$~kpc$^2$ of the central disk of NGC 628 to probe the physics of star formation.  Systematic spatial offsets between  molecular clouds and \ion{H}{2} regions illustrate the time evolution of star-forming regions. Using uniform sampling of both maps on $50{-}500$~pc scales, we infer molecular gas depletion times of $1{-}3$~Gyr, but also find the increase of scatter in the star formation relation on small scales is consistent with gas and \ion{H}{2} regions being only weakly correlated at the cloud ($50$~pc) scale. This implies a short overlap phase for molecular clouds and \ion{H}{2} regions, which we test by directly matching our catalog of $1502$ \ion{H}{2} regions and $738$ GMCs. We uncover only $74$ objects in the overlap phase, and we find depletion times ${>}1$~Gyr, significantly longer than previously reported for individual star-forming clouds in the Milky Way. Finally, we find no clear trends that relate variations in the depletion time observed on $500$~pc scales to physical drivers (metallicity, molecular and stellar mass surface density, molecular gas boundedness) on $50$~pc scales.
\end{abstract}


\section{Introduction}
The star formation relation \citep{Kennicutt1989, Kennicutt1998} identified a fundamental connection between the surface density of gas and star formation (SF). Measurements of integrated galaxies \citep{Young1996, Saintonge2011} and resolved kpc scales \citep{Kennicutt2007, Bigiel2008, Leroy2013,Meidt2015} have revealed that variations in the efficiency of star formation correlate with local physical conditions and galactic dynamics.  These conditions regulate the time evolution along the star forming sequence (molecular cloud formation, collapse, ionization of \ion{H}{2} regions, and cloud disruption), with individual stages depending strongly on local physics and physical conditions relevant on the typical ${\sim}50$~pc scales of giant molecular clouds (GMCs) and \ion{H}{2} regions.  Thus to understand the regulation of star formation it is essential to study both large scales (to understand the influence of galaxy dynamics and to time average the SF sequence) and small scales (to constrain the physics).   

The PHANGS (Physics at High Angular resolution in Nearby GalaxieS) collaboration is gathering the observations necessary to bridge these scales. With an ALMA large program we are mapping the CO emission at cloud scales across the disks of $74$ nearby galaxies (\citealt{Sun2018}, Leroy in prep). At matched resolution, our VLT/MUSE large program is mapping the ionized gas and stellar populations across a subsample of $19$ galaxies.  

We present first results from these ALMA and MUSE surveys, comparing star formation and molecular gas in NGC 628 at $50$~pc resolution. This galaxy is a nearby ($9.6$ Mpc, \citealt{Kreckel2017}) face-on ($i=9^\circ$, \citealt{Blanc2013}) type SAc grand-design spiral galaxy with a moderate star formation rate ($2.4$ M$_\sun$ yr$^{-1}$; \citealt{Sanchez2011}). We examine variations in the gas and star formation surface density ($\Sigma_{\textrm{mol}}$ and $\Sigma_{\textrm{SFR}}$) across the galaxy, parameterized by the molecular gas depletion time (\tdep = $\Sigma_{\textrm{mol}}/\Sigma_{\textrm{SFR}}$).  Using uniform sampling, we trace changes in the depletion time as a function of scale, from $50{-}500$~pc. We catalog and match individual GMCs and \ion{H}{2} regions to characterize the overlap phase in the star formation sequence. Finally, we explore which local physical conditions drive large-scale variations in the depletion time. 

\section{Data}
As the data reduction has been detailed in previous work \citep{Leroy2016, Sun2018, Kreckel2016, Kreckel2017}, we provide a brief summary here and describe the data products used. 

\subsection{Molecular Gas Tracers}
Our ALMA observations map CO \mbox{(2-1)} line emission across the central $3\arcmin \times 4\arcmin$ ($8.5 \times 11.3$ kpc$^2$) star-forming disk (Figure~\ref{fig:maps}).  They include 12m, 7m and total power observations to recover emission on all scales, achieve $1\arcsec$ (47~pc) resolution, and reach an rms noise of $0.15$~K over $2.5$~\kms\ and ($1\sigma$) sensitivity to the integrated line intensity of $1.1$~K~\kms\ ($7.5$~M$_\odot$~pc$^{-2}$) over $20$~\kms.

We assume a fixed Milky Way CO-to-H$_2$ conversion factor of $\alpha_\mathrm{CO}^{1-0}$ = 4.4 M$_\odot$~pc$^{-2}$ (K~km~s$^{-1}$)$^{-1}$ \citep{Leroy2011,Blanc2013a,Sandstrom2013}, which includes a factor of $1.36$ to account for heavy elements. We assume a \mbox{CO (2-1)/(1-0)} brightness temperature ratio of R$_{21} = 0.61$ as measured by \cite{Cormier2018}. 

$738$ GMCs are identified using CPROPS \citep{Rosolowsky2006}, an algorithm that identifies emission peaks and determines their macroscopic properties; we refer to \citet{Rosolowsky2018} for further details.  We find median cloud masses of $5.8{\times}10^5$~M$_\sun$ and median radii of $75$~pc.

\subsection{Star Formation Tracers}
We obtained MUSE observations at $1\arcsec$ seeing (Figure~\ref{fig:maps}) that achieve 0\farcs2 astrometric accuracy (calibrated off SDSS r-band imaging; \citealt{Kreckel2017}), and use LZIFU \citep{Ho2016} to simultaneously fit the stellar continuum and emission lines.  We reach a $3\sigma$ surface brightness sensitivity for H$\alpha$ of $1.5\times10^{-17}$ erg~s$^{-1}$ cm$^{-2}$~arcsec$^{-2}$. 

We correct for dust obscuration using the Balmer decrement assuming case~B recombination, an electron temperature of $10^4$~K, a \cite{Fitzpatrick1999} extinction curve  and a Milky Way value of $R_V=3.1$ for all regions. This method shows good ($\sim$5\%) systematic agreement at high $\Sigma_\textrm{SFR}$ with a hybrid H$\alpha$+24$\mu$m dust correction, suggesting no star-forming regions are completely obscured. H$\alpha$ is detected across $95\%$ of the map, however H$\beta$ is only detected in ${\sim}50\%$ of pixels (corresponding to morphologically diffuse H$\alpha$ emission).  We assume a fixed $A_V=1.3$ mag where H$\beta$ is not detected, consistent with typical \ion{H}{2} region values ($0.5{-}1.5$ mag) and $\sim$1~kpc integrated diffuse regions. We note that our results are not sensitive to this choice as most of this more diffuse emission is subtracted from our SFR maps (Section~\ref{sec:dig}).

We construct a catalog of $1502$ \ion{H}{2} regions using \mbox{HIIPhot} \citep{Thilker2000}. We detect minimum H$\alpha$ luminosities of $3 \times 10^{36}$ erg~s$^{-1}$, with median luminosities of $10^{37}$ erg~s$^{-1}$ (comparable to the ionizing flux produced by a single O8V star; \citealt{Schaerer1997}) and radii of $35$~pc (only marginally resolved).  For each we calculate the gas-phase oxygen abundance using the theoretical strong line N2S2 diagnostic of \cite{Dopita2016}.

We convert our H$\alpha$ luminosities into star formation rate (SFR) estimates following \cite{Murphy2011} as 
\begin{equation}
\mathrm{SFR} (\mathrm{M}_\odot\,\mathrm{yr}^{-1}) = 5.37 \times 10^{-42} \mathrm{L}({\rm H}\alpha) (\mathrm{erg}\,\mathrm{s}^{-1})\,.
\end{equation}
We caution that on the scale of individual regions likely dominated by a single stellar population, the concept of a continuous star formation rate begins to break down \citep[potentially leading to systematic underestimation of SFR by a factor of $2{-}3$;][]{Faesi2014}. 

\subsection{Treatment of diffuse gas}
\label{sec:dig}

The treatment of ``diffuse'' CO emission is an open, unsolved issue.  While the existence of faint, spatially extended 12CO emission in some galaxies has been clearly established \citep{Pety2013,Roman-Duval2016}, the detailed structure and physical properties (e.g. density) of this gas -- and its dependence on properties of the host galaxy -- is less clear.  In external galaxies, one cannot easily distinguish between a truly diffuse CO-emitting molecular phase and a uniform distribution of small clouds that are separated by less than the beam size.  As the goal of this analysis is to compare the star formation with the available molecular gas reservoir, we do not attempt to remove a diffuse CO gas component from our data.

As shown in \cite{Kreckel2016}, a significant fraction (${\sim}20{-}50\%$) of the H$\alpha$ emission in NGC 628 arises from a diffuse ionized gas (DIG) component, but its association with star formation is not clear \citep{Zurita2000,Zhang2017}. We choose here to model and subtract the DIG emission from our star formation calculations, with the caveat that we could be systematically underestimating the SFR on large scales. 

We leverage the [\ion{S}{2}]/H$\alpha$ $>$ 0.5 line ratio to identify 10\% of the total H$\alpha$ flux (41\% of the pixels) as pure DIG emission in our extinction corrected H$\alpha$ line map \citep{Kreckel2016}. We interpolate across the entire map, and subtract this DIG model from our extinction corrected H$\alpha$ map (Figure~\ref{fig:maps}).  This removes $16\%$ of the flux at the location of cataloged \ion{H}{2} regions, and results in a total diffuse fraction of $36\%$.

\section{Results}

\subsection{Scatter in the SF Relation - variations with scale} 
The tight correlation between SFR surface density and molecular gas surface density observed on $>$kpc scales in galaxies \citep{Bigiel2008, Bigiel2011, Leroy2013,Utomo2017}  breaks down on smaller scales due to stochastic sampling of the evolution of individual  molecular clouds and SF regions \citep{Schruba2010, Onodera2010, Kruijssen2014, Jameson2016}.  This is apparent in NGC 628 (Figure~\ref{fig:maps}) by the clear displacement of the H$\alpha$ and CO emission due to time evolution combined with the motion of the spiral pattern around the disk \citep{Schinnerer2017}.  Figure~\ref{fig:scls} explores how the inferred depletion times and their scatter change over $50{-}500$~pc scales.  Individual data points show the surface density in (Nyquist sampled) apertures detected in both tracers uniformly sampling the emission maps at $\textrm{S/N}>3$.  The inset histograms show that a significant number of regions (50-75\%) are detected in only one tracer. 

We measure median gas depletion times of $1{-}3$ Gyr, consistent with what was previously found for NGC 628 on ${\sim}400$~pc \citep{Rebolledo2015} and $750$~pc \citep{Bigiel2008, Leroy2008, Leroy2013} scales.  At scales below $300$~pc the molecular gas and SFR surface densities are largely uncorrelated (Spearman's rank correlation coefficient $\rho \approx 0.3$), and even on larger scales the correlation remains weak ($\rho \approx 0.5$).  This weak correlation combined with the large number of regions detected in only one tracer implies there is only a short overlap phase between molecular gas and young stars.

We observe an increased scatter in the depletion time ($\sigma \approx 0.4$ dex) on small (${<}300$~pc) scales, consistent with what has been observed in M51 \citep{Blanc2009}, M33 \citep{Schruba2010}, and the LMC and SMC \citep{Jameson2016}.  By combining our results on small scales with complementary measurements for NGC 628 from the HERACLES survey \citep{Leroy2013}, we can trace the change in scatter from $50$~pc to $2.4$~kpc scales (Figure~\ref{fig:scls2}), observing a flattening in the scatter below ${\sim}300$~pc.  

If neighboring star-forming regions are independent and uncorrelated, with evolutionary cycling as the only source of scatter, then we naively expect this relation to follow a simple power law ($\sigma \propto R^{-1}$; \citealt{Leroy2013}). However, accounting for additional scatter from the cloud or \ion{H}{2} region mass spectrum, flux evolution within a single evolutionary phase, and sensitivity limits, we expect uncorrelated star formation to result in a flattening at the smallest spatial scales \citep{Kruijssen2014, Kruijssen18}. Similarly, the smallest apertures contain on average only a single GMC and \ion{H}{2} region, which causes the scatter to decrease. The observed trend agrees qualitatively with the predictions of the \cite{Kruijssen2014} model (Figure~\ref{fig:scls2}), where for reference we show a curve of the model assuming standard parameter values, specifically measured within NGC 628 when possible \citep{Chevance2018}.

\subsection{Direct cloud scale constraints and comparison to the Milky Way and Local Group Galaxies}

Given our large sample of $738$ GMCs and $1502$ \ion{H}{2} regions, we expect some fraction of these will currently exist in a short overlap phase \citep{Kawamura2009}.  To compare with cloud based studies in the Milky Way and Local Group galaxies, we cross-match all GMC and \ion{H}{2} region centers (crossmatch tolerance of $0\farcs5 \approx 20$~pc) 
and find $74$ matches (Figure~\ref{fig:tot2}, left).  
By eye, ${>}90\%$ of these look like associated objects, and not chance alignments in crowded regions.  
The small fraction of overlapping regions implies there is a quick time evolution between molecular gas clouds and young stars.

We infer integrated cloud scale depletion times (M$_\textrm{GMC}$/SFR) of $1{-}10$ Gyr, with a median value ($3$~Gyr) slightly longer than the depletion times we measure over larger scales (Figure~\ref{fig:scls}).  As each GMC may collapse to form multiple stars and star clusters, all additional \ion{H}{2} regions within the GMC footprint are also included in calculating the associated SFR.  We note that GMCs that are not fully decomposed could bias us to longer depletion times. 

As the typical GMC size is larger than the typical \ion{H}{2} region ($75$ vs $35$~pc in radius, Figure \ref{fig:maps}), we perform a second catalog match by loosening the requirement that the selected GMCs must have HII region counterparts within 0$\farcs$5 and instead sum up all \ion{H}{2} regions where the center falls within a given GMC's footprint.  By selecting for peaks in the molecular gas distribution and measuring the associated star formation rate, an approach that biases us to longer depletion times ($\sim$10~Gyr), we find $162$ GMCs with associated star formation (Figure~\ref{fig:tot2}, right), of which $40\%$ have two or more \ion{H}{2} regions associated.  

Milky Way studies \citep{Lada2010, Murray2011, Evans2014, Vutisalchavakul2016} show systematically shorter ($100$~Myr to $1$~Gyr) depletion times (Figure~\ref{fig:tot2}), with over two orders of magnitude scatter (partly driven by methodological differences between studies). We find good agreement when extrapolating the trend identified by \cite{Vutisalchavakul2016}, though our clouds are an order of magnitude more massive.  Few such cloud scale studies are available for external galaxies.  \cite{Ochsendorf2017} observed long (${\sim}1$ Gyr) depletion times as a function of GMC mass in the LMC.  Targeting \ion{H}{2} regions in NGC 300, \cite{Faesi2014} measured significantly short ($230$~Myr) depletion times on $250$~pc scales.  We note that studies preselecting only star-forming regions biases their results to shorter depletion times \citep{Kruijssen18}, just as our preselection of GMCs (Figure~\ref{fig:tot2}, right) biases us to longer depletion times.

\subsection{Connecting \tdep\ with physical conditions on 50pc scales}

We observe ${\sim}0.3$ dex scatter in \tdep\ on large scales. In M51, \cite{Meidt2013} and \cite{Leroy2017} found such variations to correlate with the apparent gravitational boundedness of the gas, while theories focused on the gravitational free-fall time predict correlations between \tdep\ and the local mean cloud density. We test for such variations by calculating \tdep\ at $500$~pc scales and then correlating this with the local mass-weighted mean properties of the molecular gas. This methodology \citep{Leroy2016} tests how the mean cloud-scale properties affect the depletion time.

In Figure~\ref{fig:pixprops2} we consider trends with the molecular gas, SFR and stellar surface densities. The latter is modeled from Spitzer imaging and cleaned of non-stellar emission \citep{Querejeta2015}.  We also explore the impact of local physical conditions, including the molecular gas boundedness ($B \equiv \Sigma_\textrm{mol} / \sigma_v^2$, where $\sigma_v$ is the line equivalent width), gas phase metallicity measured within \ion{H}{2} regions, and \ion{H}{2} region clustering (parameterized by $n_3$, the average projected distance to the three nearest neighbors).  

We observe the strongest correlation with SFR surface density, however this parameter covers a wider dynamic range than the molecular gas surface density (which shows no correlation). Random sampling of the CO surface densities does not change the strength of this correlation, 
suggesting it is purely driven through the inverse relationship with \tdep.    The central star-forming ring (at high stellar mass surface densities) shows systematically longer depletion times ($4{-}8$ Gyr), though the reported decrease in CO-to-H$_2$ conversion factor in the center implies an overestimation by up to a factor of two \citep{Blanc2013a, Sandstrom2013}, but cannot fully account for the offset we see.  Gas that is more bound (larger~$B$) was found in M51 to exhibit shorter depletion time \citep{Leroy2017}, however we observe no such correlation in NGC 628.  This suggests the key driver for long depletion times in M51 is not simply the virial parameter of the molecular gas, but large-scale dynamical effects as suggested by \cite{Meidt2013}.  Other trends in NGC 628 are only tentative ($\rho \leq 0.3$), but suggest shorter depletion times occur in the outer disk (lower stellar mass surface densities) where \ion{H}{2} regions are more clustered (lower $n_3$). A~larger sample of galaxies is needed to expand the parameter space explored here.

\section{Implications}

\begin{figure*}
\begin{flushright}
\includegraphics[width=6.3in]{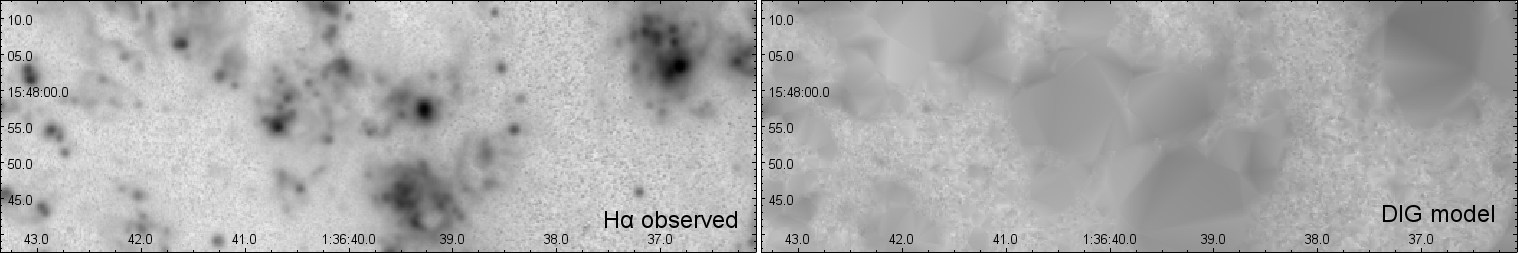}
\end{flushright}
\centering
\includegraphics[width=7in]{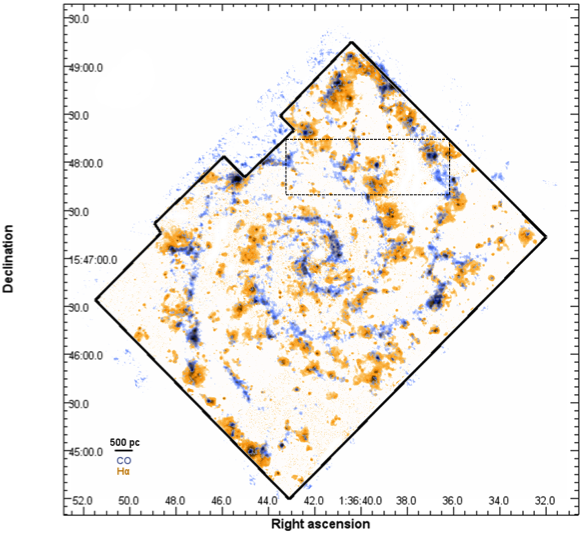}
\begin{flushright}
\includegraphics[width=6.3in]{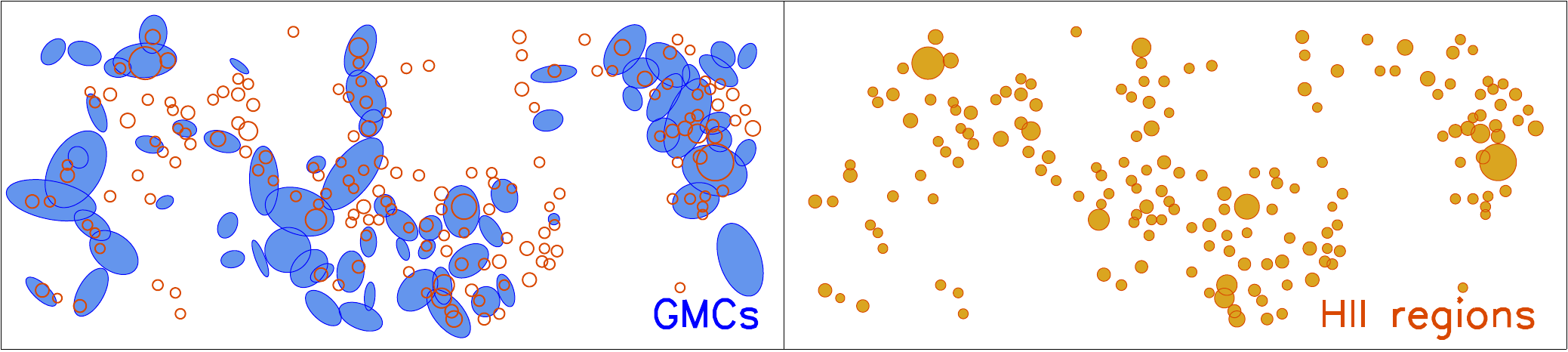}
\end{flushright}
\caption{Our ALMA CO \mbox{(2-1)} (blue) and MUSE H$\alpha$ (orange) intensity maps cover the central $8.5 \times 11.3$ kpc$^2$ star-forming disk of NGC 628 at $47$~pc resolution (middle).  GMCs and \ion{H}{2} regions are clearly resolved into discrete structures.  A zoom-in region highlights the diffuse H$\alpha$ emission surrounding compact \ion{H}{2} regions (top left), which we model (top right) and subtract, and demonstrates with simplified ellipses (bottom) the cataloged GMCs (blue) and HII regions (orange).
\label{fig:maps}}
\end{figure*}

\begin{figure*}
\centering
\includegraphics[width=7in]{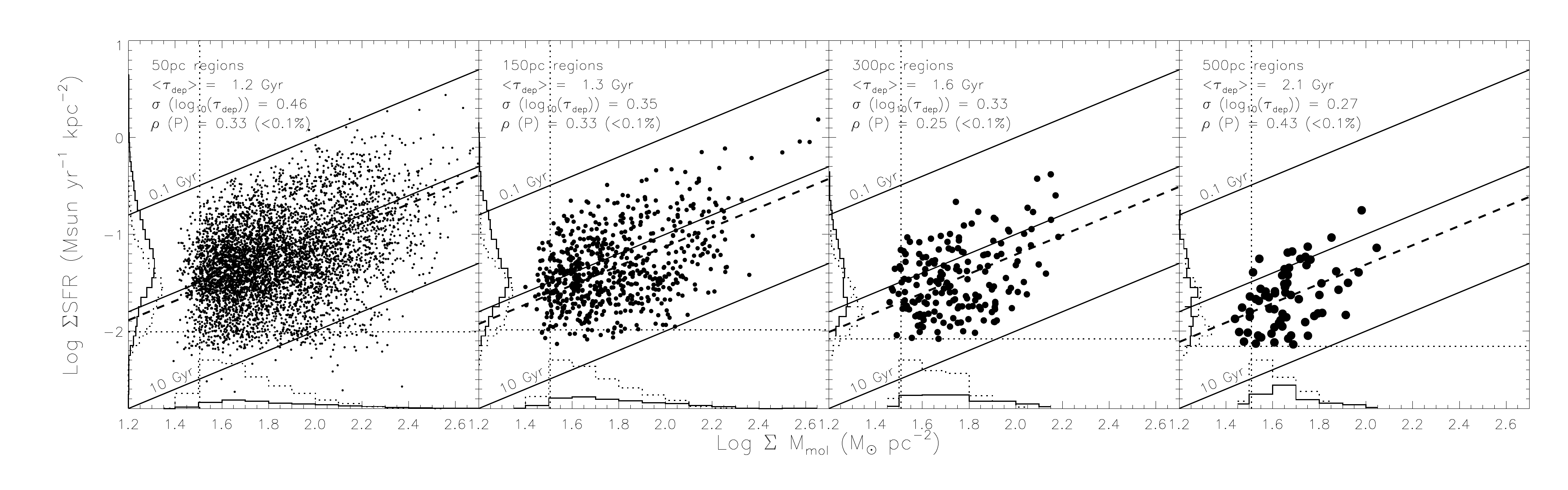}
\caption{Variations in SFR surface density ($\Sigma_\textrm{SFR}$) as a function of the molecular gas surface density ($\Sigma_\textrm{mol}$) on spatial scales ranging from $50$~pc to $500$~pc.  The median inferred depletion time (\tdep), scatter and Spearman's rank correlation coefficient ($\rho$) and its significance ($P$) are shown.  Constant depletion times (solid lines) and the median value (dashed line) are overplotted. Dotted lines show the $3\sigma$ sensitivity limits in each tracer. The distribution of regions detected in both tracers (solid) or only one (dotted) are shown as inset histograms. Star formation and molecular gas have substantial scatter on all scales. 
\label{fig:scls}}
\end{figure*}

\begin{figure*}
\centering
\includegraphics[width=3in]{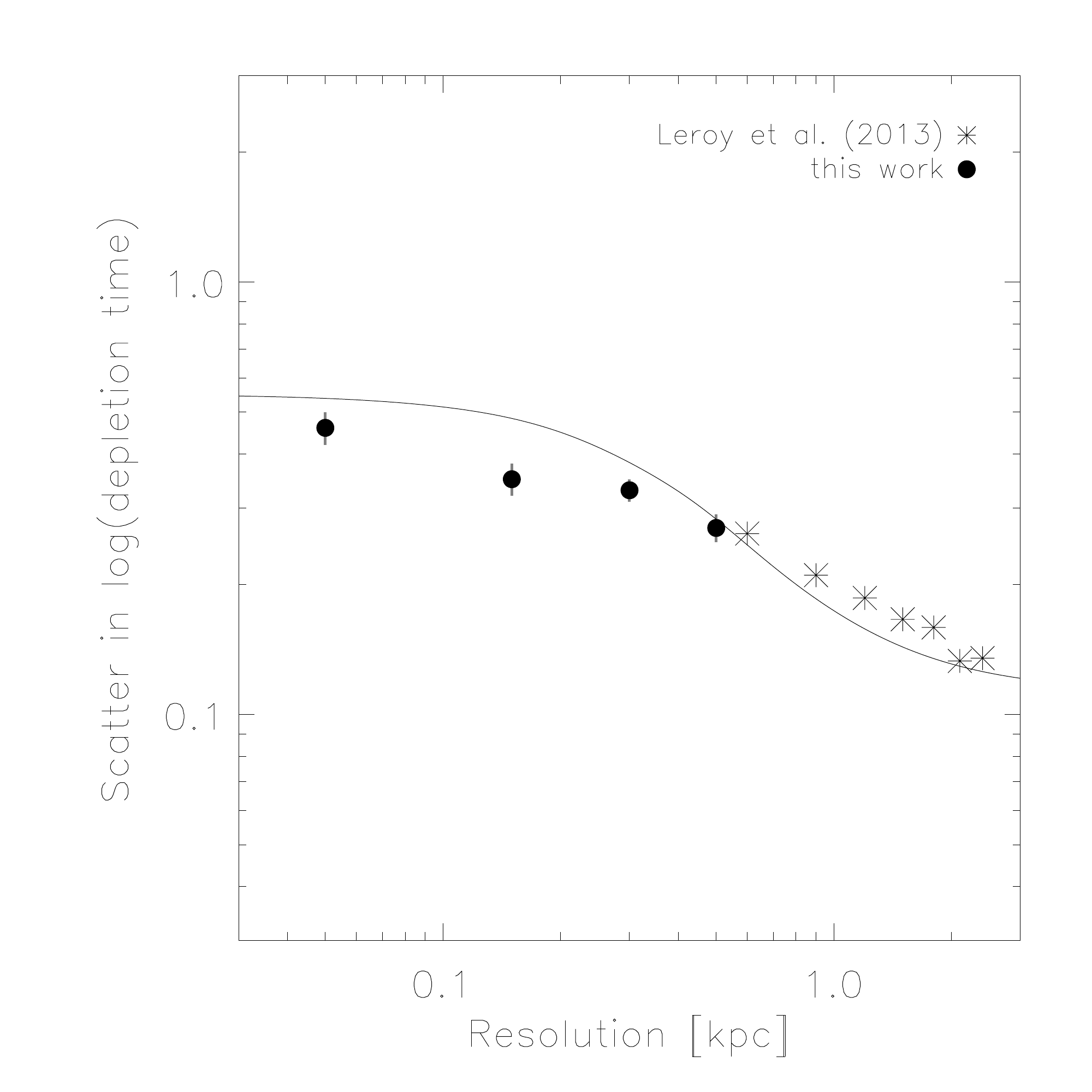}
\caption{Scatter in the depletion time as a function of scale, combining our $50$~pc to $500$~pc results (circles) with \cite{Leroy2013} at $600$~pc to $2.4$~kpc scales.  Following \cite{Kruijssen2014}, detailed modeling of the expected relation assuming uncorrelated neighboring regions (solid line) shows relatively good agreement and reproduces the flattening at small scales. Here we have adopted the following model parameters: H$\alpha$ lifetime of $5$~Myr \citep{Haydon2018}, GMC lifetime of $25$~Myr, overlap timescale of $3$~Myr, characteristic separation of $125$~pc, equal flux ratio between isolated emission and the emission in overlap regions, a scatter due to luminosity evolution of the stars and gas of $0.3$~dex, a scatter due to the molecular cloud mass spectrum of $0.33$~dex \citep{Rosolowsky2018}, and a scatter due to observational uncertainties of $0.1$~dex.  Many of these parameters are degenerate, however we take here the best estimates currently available, and specifically measured within NGC 628 when possible \citep{Chevance2018}.
\label{fig:scls2}}
\end{figure*}

Although star formation and molecular gas are organized into similar structures in NGC 628 (Figure~\ref{fig:maps}), physical offsets of more than $100$~pc are apparent and uniform sampling reveals large variations in depletion time on all scales ($50{-}500$~pc, Figure~\ref{fig:scls}).  As the scatter among the local depletion times is well modeled by star formation in neighboring regions being uncorrelated (Figure~\ref{fig:scls2}), the rotation of the spiral pattern organizing and concentrating the molecular gas combined with the time evolution of star formation regions result in GMCs and \ion{H}{2} regions being only weakly correlated at cloud scales. 

\begin{figure*}
\centering
\includegraphics[width=3.5in]{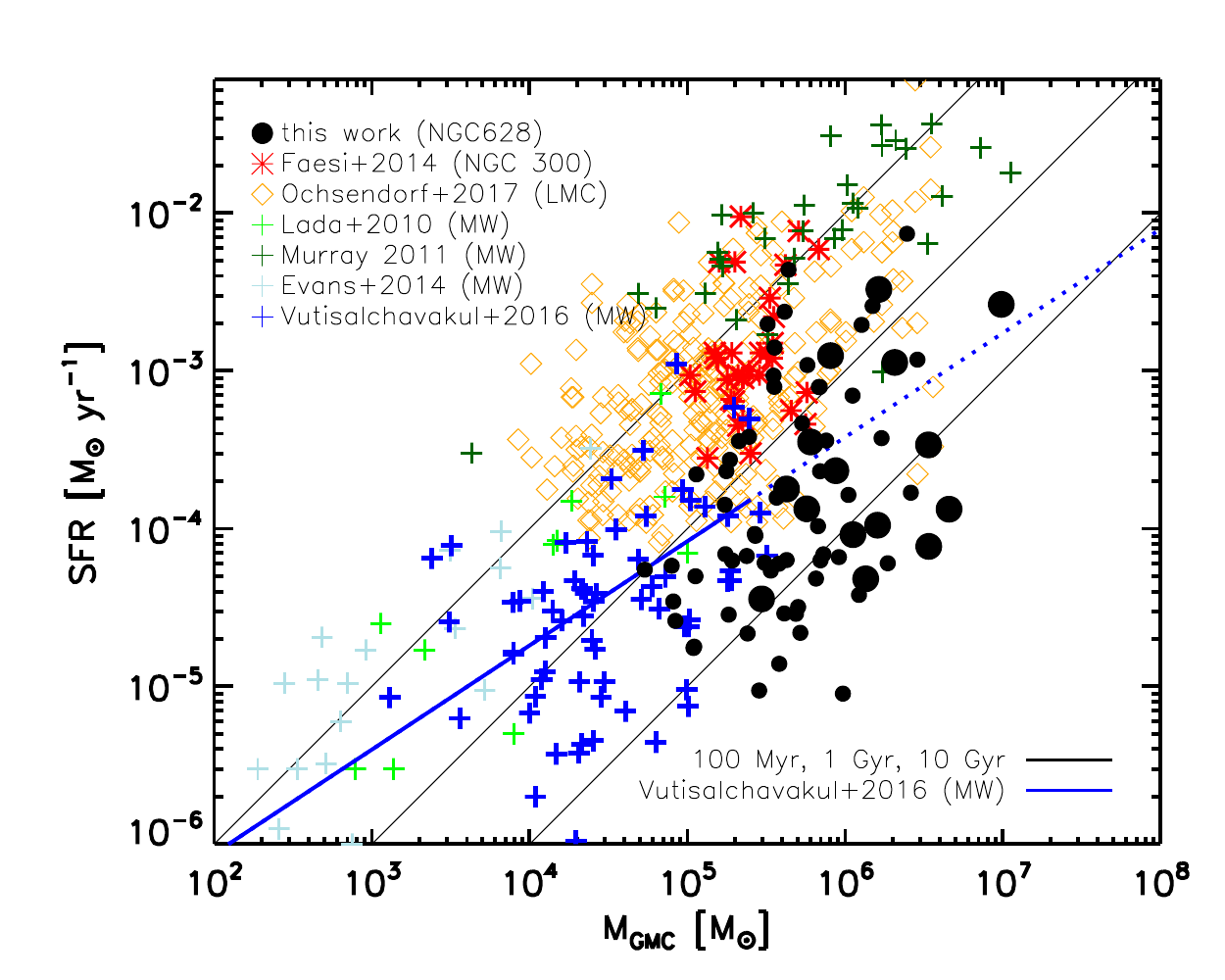}
\includegraphics[width=3.5in]{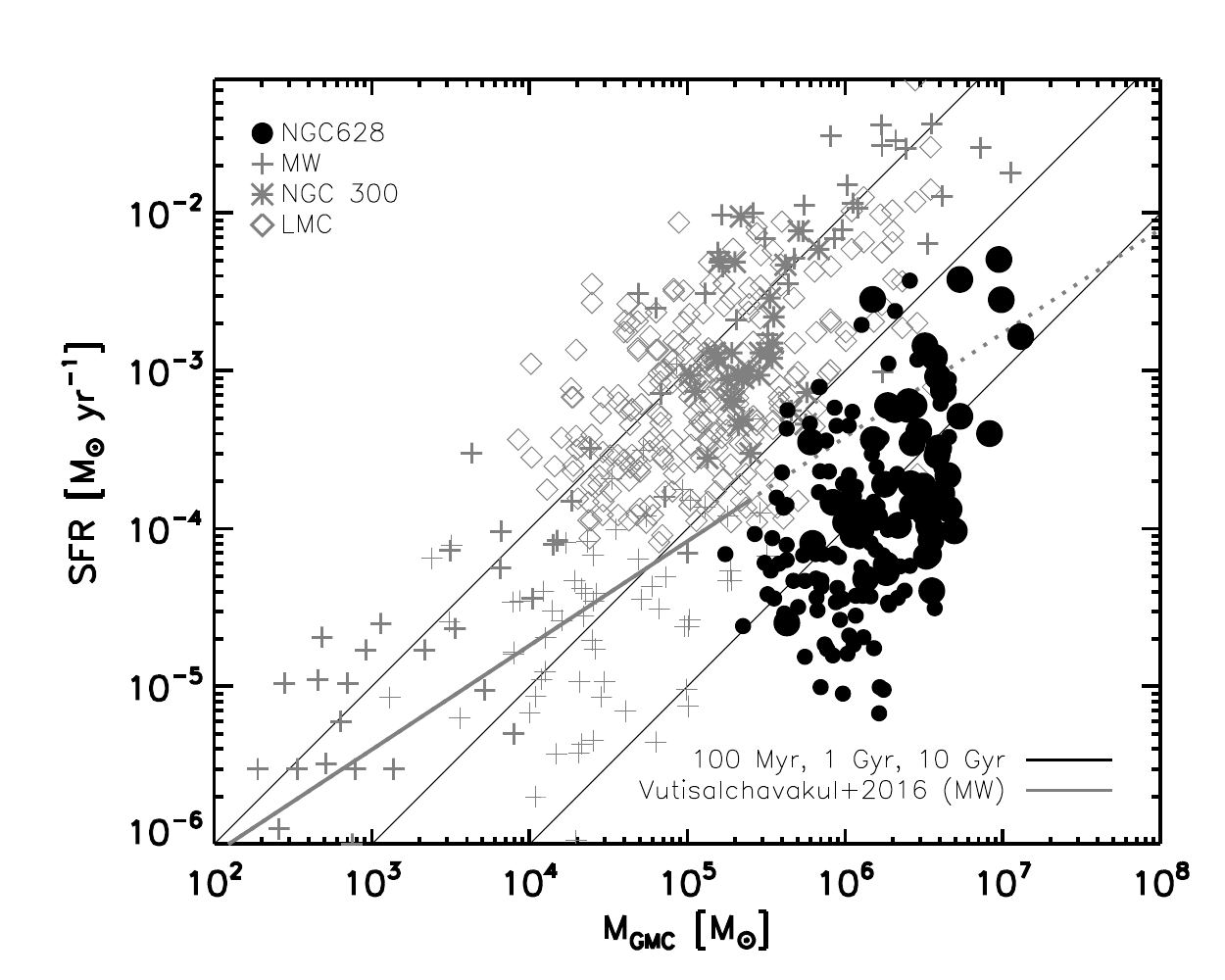}
\caption{Star formation rate as a function of GMC mass using two different methods of matching GMC and HII region catalogs: overlapping centers (left) or summing all \ion{H}{2} regions within a given GMC footprint (right).  A larger symbol size indicates if that GMC has more than one associated \ion{H}{2} region.  We compare to literature results (see text) and the Milky Way relation (extrapolation shown as a dotted line; \citealt{Vutisalchavakul2016}).  We recover a wider range of depletion times and little correlation between the two tracers, unlike in previous studies, but consistent with the expectations for scatter driven by evolutionary cycling. 
\label{fig:tot2}}
\end{figure*}

Our wide map provides the statistics to identify $74$ spatially coincident objects existing in a short overlap phase out of $738$ GMCs and $1502$ \ion{H}{2} regions. These exhibit longer depletion times ($1{-}10$ Gyr) than individual Milky Way clouds (Figure~\ref{fig:tot2}). Even if the SFR of individual regions is systematically underestimated by a factor of $2{-}3$ (based on DIG subtraction and lack of SF history modeling), and our GMCs are not fully decomposed, this cannot account for the discrepancy with previous results.  As ${\sim}40\%$ of GMCs with associated star formation overlap with multiple \ion{H}{2} regions, it is clear that there is a non-trivial connection between GMC mass function and \ion{H}{2} region luminosity function. 

\begin{figure*}
\centering
\includegraphics[width=7in]{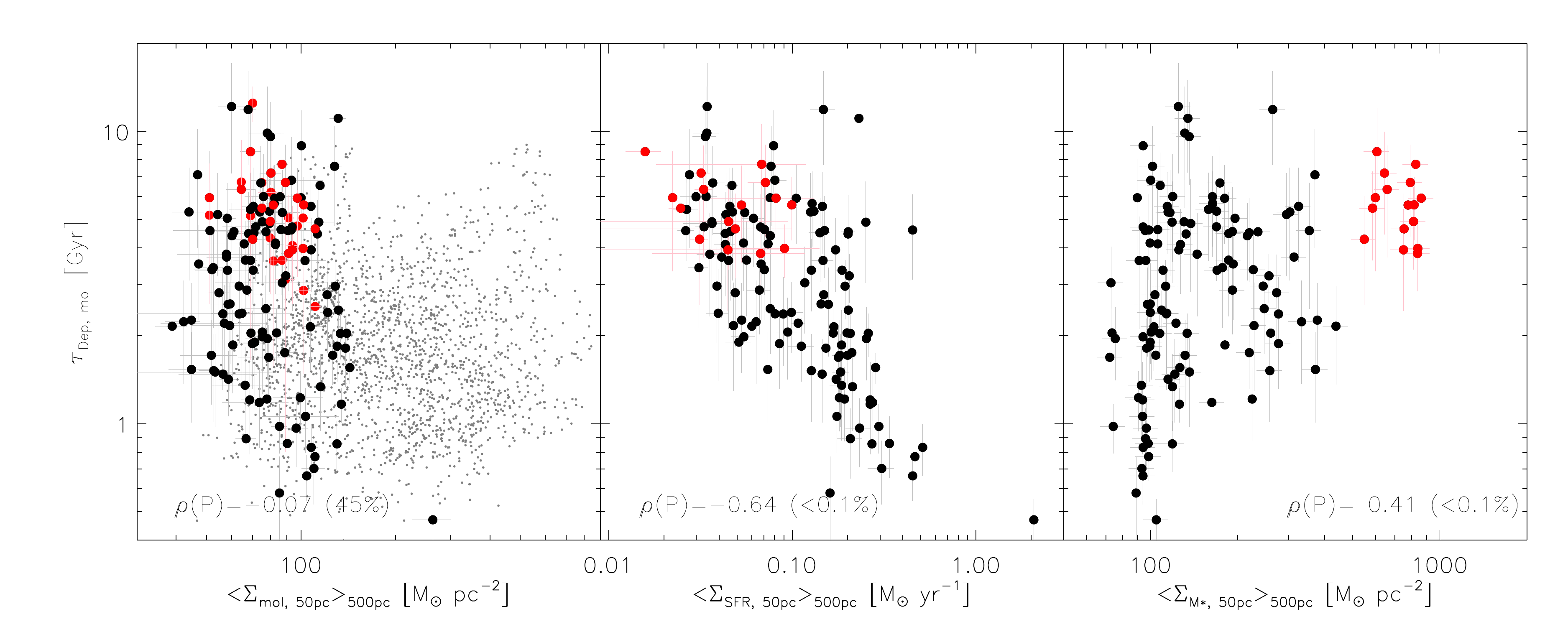}
\includegraphics[width=7in]{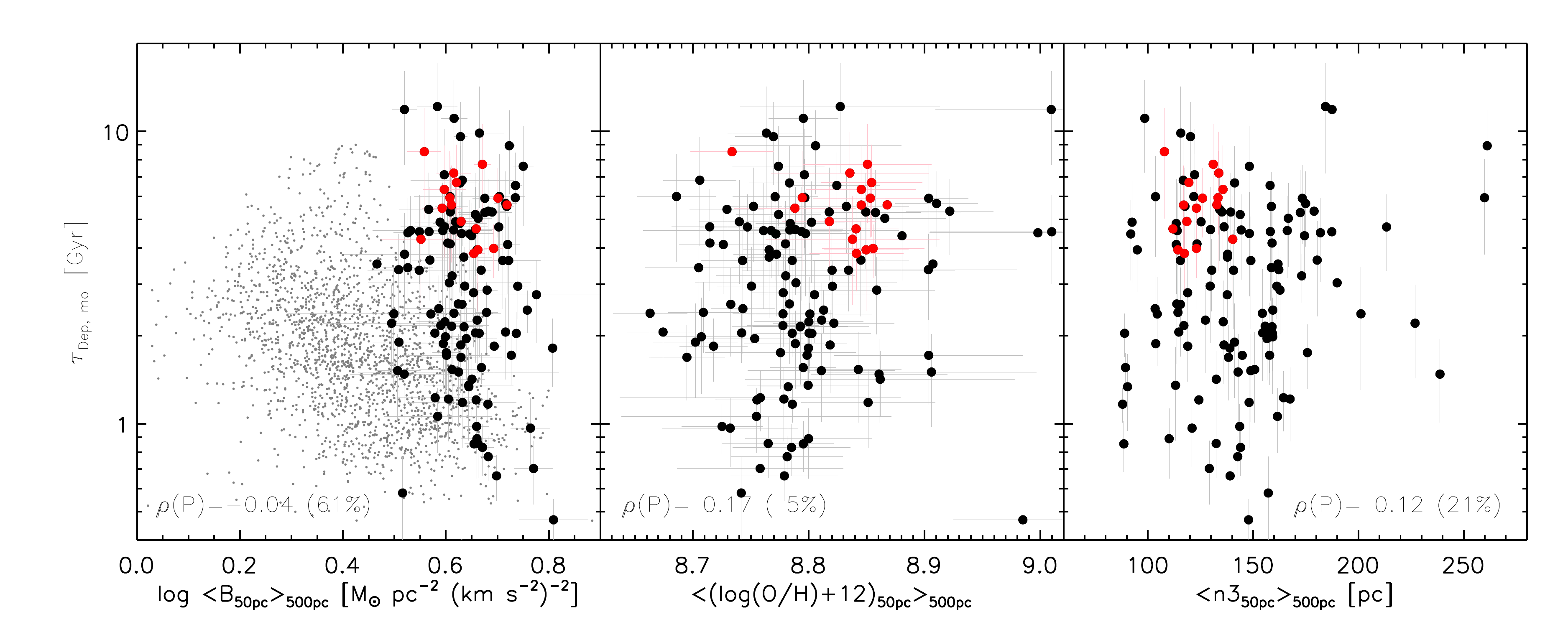}
\caption{Variations in the depletion time (\tdep) at $500$~pc scale as a function of the surface density (top, left to right) of different physical parameters (see text). These properties reflect the luminosity weighted average value observed on $50$~pc scales within $500$~pc regions, with Spearman's rank correlation coefficient ($\rho$) and significance ($P$) shown. Results for M51 are shown in grey \citep{Leroy2017}. We observe the strongest correlation with the SFR surface density, however this may be driven by correlated axes and the larger dynamic range probed in this tracer compared to the molecular gas surface density.  The central star-forming ring (red) also shows systematically longer depletion times.
\label{fig:pixprops2}}
\end{figure*}

The physical parameters we investigate (gas boundedness, metallicity, \ion{H}{2} region clustering, and molecular gas, SFR and stellar surface density) do not strongly drive the large-scale variations in \tdep\ that we observe across the disk (Figure~\ref{fig:pixprops2}).  Our results show only suggestive trends, with shorter depletion times occurring in the outer disk ($>$5~kpc) where \ion{H}{2} regions are more clustered and the surface density of star formation is higher. Alternately, this could suggest that variations in \tdep\ on 500 pc scales are predominantly driven by dynamical effects on even larger scales.  We do observe systematically longer depletion times in the central star-forming ring, where variations in the CO-to-H$_2$ conversion factor account for only some of the offset \citep{Blanc2013a,Sandstrom2013}, suggesting the dynamical environment plays an important role. 

Isolating the local physical conditions that drive changes in the star formation efficiency requires a more systematic sampling of the parameter space, beyond what is possible in this first case study. 

\acknowledgments
We thank the referee for their helpful comments that improved this work.  KK gratefully acknowledges support from grant KR 4598/1-2 from the DFG Priority Program 1573.  
JMDK and MC gratefully acknowledge funding from the German Research Foundation (DFG) in the form of an Emmy Noether Research Group (grant number KR4801/1-1). JMDK gratefully acknowledges funding from the European Research Council (ERC) under the European Union's Horizon 2020 research and innovation programme via the ERC Starting Grant MUSTANG (grant agreement number 714907). BG gratefully acknowledges the support of the Australian Research Council as the recipient of a Future Fellowship (FT140101202).
FB acknowledges funding from the European Union's Horizon 2020 research and innovation programme (grant agreement No 726384 - EMPIRE). 
GB is supported by CONICYT/FONDECYT, Programa de Iniciaci\'on, Folio 11150220. 
AH acknowledges support from the Centre National d'Etudes Spatiales (CNES). 
ER acknowledges the support of the Natural Sciences and Engineering Research Council of Canada (NSERC), funding reference number RGPIN-2017-03987. 
RM and ES acknowledge funding from the European Research Council (ERC) under the European Union's Horizon 2020 research and innovation programme (grant agreement No. 694343). JP acknowledges support by the Programme National "Physique et Chimie du Milieu Interstellaire"(PCMI) of CNRS/INSU with INC/INP co-funded by CEA and CNES.

Based on observations collected at the European Organisation for Astronomical Research in the Southern Hemisphere under ESO programme 094.C-0623, ID 095.C-0473 and  098.C-0484.

This paper makes use of the following ALMA data: ADS/JAO.ALMA\#2012.0.00650.S. ALMA is a partnership of ESO (representing its member states), NSF (USA) and NINS (Japan), together with NRC (Canada), MOST and ASIAA (Taiwan), and KASI (Republic of Korea), in cooperation with the Republic of Chile. The Joint ALMA Observatory is operated by ESO, AUI/NRAO and NAOJ. The National Radio Astronomy Observatory is a facility of the National Science Foundation operated under cooperative agreement by Associated Universities, Inc.

\bibliographystyle{yahapj}

\clearpage


\end{document}